\documentclass[equations,11pt]{article}
\usepackage{amssymb,amsmath,amsthm}
\usepackage{eucal}

\renewcommand{\theequation}{\arabic{equation}}
\begin{document}
\bibliographystyle{plain}
\def\m@th{\mathsurround=0pt}
\mathchardef\bracell="0365 
\def\upbrall{$\m@th\bracell$}
\def\undertilde#1{\mathop{\vtop{\ialign{##\crcr
    $\hfil\displaystyle{#1}\hfil$\crcr
     \noalign
     {\kern1.5pt\nointerlineskip}
     \upbrall\crcr\noalign{\kern1pt
   }}}}\limits}
\def\theequation{\arabic{section}.\arabic{equation}}
\newcommand{\pp}{\partial}
\newcommand{\ar}{\alpha}
\newcommand{\aar}{\bar{a}}
\newcommand{\bb}{\beta}
\newcommand{\gm}{\gamma}
\newcommand{\Gm}{\Gamma}
\newcommand{\en}{\epsilon}
\newcommand{\ven}{\varepsilon}
\newcommand{\dd}{\delta}
\newcommand{\sg}{\sigma}
\newcommand{\kp}{\kappa}
\newcommand{\ld}{\lambda}
\newcommand{\oa}{\omega}
\newcommand{\hf}{\frac{1}{2}}
\newcommand{\be}{\begin{equation}}
\newcommand{\ee}{\end{equation}}
\newcommand{\bea}{\begin{eqnarray}}
\newcommand{\eea}{\end{eqnarray}}
\newcommand{\bse}{\begin{subequations}}
\newcommand{\ese}{\end{subequations}}
\newcommand{\nn}{\nonumber}
\newcommand{\bR}{\bar{R}}
\newcommand{\bP}{\bar{\Phi}}
\newcommand{\bS}{\bar{S}}
\newcommand{\bu}{{\boldsymbol u}}
\newcommand{\bt}{{\boldsymbol t}}
\newcommand{\bm}{{\boldsymbol m}}
\newcommand{\boa}{{\boldsymbol \omega}}
\newcommand{\bet}{{\boldsymbol \eta}}
\newcommand{\bW}{\bar{W}}
\newcommand{\vf}{\varphi}
\newcommand{\sn}{{\rm sn}}
\newcommand{\wh}{\widehat}
\newcommand{\ol}{\overline}
\newcommand{\wt}{\widetilde}
\newcommand{\ut}{\undertilde}
\newtheorem{theorem}{Theorem}[section]
\newtheorem{lemma}{Lemma}[section]
\newtheorem{cor}{Corollary}[section]
\newtheorem{prop}{Proposition}[section]
\newtheorem{definition}{Definition}[section]
\newtheorem{conj}{Conjecture}[section]

\phantom{\vspace{6cm}} 

\begin{center} 
{\Large{\bf Discrete Dubrovin Equations and \\ 
Separation of Variables for Discrete Systems } 
\vspace{.8cm}

Frank W. Nijhoff}  \vspace{.2cm} \\ 
{\it Department of Applied Mathematics\\
The University of Leeds, Leeds LS2 9JT, UK}\\
\vspace{.4cm}
June 9, 1998 
\end{center} 
\vspace{4cm}

\vfill\noindent 
To appear in: Proceedings of the International Conference on 
{\it Integrability and Chaos in Discrete Systems} 
Brussels, July 2-6, 1997.

\pagebreak 

\setcounter{page}{0}

\begin{center} 
{\Large{\bf Discrete Dubrovin Equations and \\ 
Separation of Variables for Discrete Systems }}
\vspace{.3cm}

F.W. Nijhoff\vspace{.2cm} \\ 
Department of Applied Mathematical Studies\\
The University of Leeds, Leeds LS2 9JT, UK
\end{center} 
\vspace{.3cm}
\centerline{\bf Abstract}
\vspace{.2cm}

A universal system of difference equations associated with a 
hyperelliptic curve is derived constituting the 
discrete analogue of the Dubrovin equations arising in the theory 
of finite-gap integration. The parametrisation of the solutions in 
terms of Abelian functions of Kleinian type (i.e. the higher-genus 
analogues of the Weierstrass elliptic functions) is discussed as well 
as the connections with the method of separation of variables. 
\vskip 2cm

\section{ Introduction} 
\setcounter{equation}{0}
The method of finite-gap integration has been developed in the late 
seventies to deal with the periodic solutions of integrable nonlinear 
evolution equations, cf. e.g. \cite{DMN,Krich,Dubr1}, and the recent 
monograph \cite{BBEIM}. 
The Dubrovin equations arise in the theory of finite-gap integration, 
namely as the equations governing the dynamics of the so-called 
auxiliary spectrum or equivalently of the poles of the Baker-Akhiezer 
function. In the case of integrable models associated with a hyperelliptic 
curve they are typically given by a first-order system of coupled ordinary 
differential equations the form, \cite{Dubr2}, 
\begin{equation} \label{eq:cDubr}
\dot{\mu}_i = \frac{\sqrt{R(\mu_i)}}{\prod_{j\neq i} (\mu_i -\mu_j)}
\  , 
\end{equation} 
in which $R(\mu)$ denotes the discriminant of the hyperelliptic 
curve in question. 
Variations on eqs. (\ref{eq:cDubr}) might occur depending on 
details of the model under consideration, but nonetheless  
the equations are to a great extent universal among the various 
integrable systems. They can generally be resolved by using the 
Lagrange interpolation formula, thus leading to the formulation of a 
Jacobi inversion problem from which the dynamics of the model 
can be solved in terms of the zeroes of theta-functions associated with 
the curve. 

In the last few years a gradual shift of emphasis has taken place 
from continuous to discrete integrable systems. One particular class 
of such discrete systems represent dynamical systems evolving 
with discrete time, namely integrable mappings, cf. 
e.g. \cite{Ves} for a review. Such mappings can arise as 
periodic solutions of partial difference analogues of equations of 
KdV type, cf. \cite{PNC,CNP}. Since these are the discrete analogues 
of the finite-gap potentials in the continuous theory, it is natural 
to expect that analogous methods can be applied to these discrete 
models as the ones used to deal with periodic problems in the 
continuous case. In a recent paper \cite{Kent} the finite-gap 
integration of 
such mappings was considered and the interpretation of the maps 
was given as constituting special addition formulae for hyperelliptic 
Abelian functions involving special winding vectors on the spectral 
curve. As a byproduct difference analogues of the Dubrovin equations 
(\ref{eq:cDubr}) were derived which form the equations of the discrete 
motion of the auxiliary spectrum under the KdV mappings. The term 
discrete Dubrovin equations stems from the paper \cite{DMN} where 
related equations were discussed but no explicit formulae were 
given. 

It is well-known that in problems associated with hyperelliptic 
curves, such as the periodic problems for equations of KdV type, the 
poles of the Baker-Akhiezer function play the role of {\em 
separation variables}, cf. e.g. \cite{FM,AvM}. Recently, this 
observation 
was cast into a wider framework by Sklyanin in \cite{Skly} using the 
Lax pair approach to obtain separation of variables for a wide class 
of integrable models. In the last section of this note we will 
comment on the fine interplay between the discrete-time integrable 
systems which are the mappings of KdV type and the separation 
mechanism.  We believe that the problem of finding the 
structure behind the discrete Dubrovin equations might form one of 
the keys to understand the separation mechanism in the new approach.

\section{Discrete Dubrovin Equations}
\setcounter{equation}{0}

In this note we investigate a class of discrete integrable systems that 
admit the following Lax description. Let us introduce the elementary 
matrices depending on a discrete variable $n$ labelling the sites along 
a chain of length $N$, given by  
\begin{equation}\label{eq:VW}
V_n(\ld) = \left(\begin{array}{cc} v_n&1\\ \ld_n&0 \end{array}\right)\   , 
\end{equation} 
in which $\ld_n=\ld+\ar_n$, $\ld$ being the spectral parameter 
and the $\ar_n$ being some site-dependent shifts. All variables 
can be taken to be real or complex valued as we wish. 
The most important object for us is the 
monodromy matrix 
\begin{equation}\label{eq:T} T(\ld) = 
\stackrel{\longleftarrow}{\prod_{n=1}^N}\,V_n(\ld) = 
\left( \begin{array}{cc} 
A(\ld) & B(\ld) \\ 
C(\ld) & D(\ld) 
\end{array} \right)\  ,   \end{equation} 
which we assume to evolve under a discrete-time map 
$v_n\mapsto \wt{v}_n$ according to 
\begin{equation}\label{eq:Tmap} 
\wt{T}(\ld)=M(\ld)T(\ld)M(\ld)^{-1}\     \ ,\     \ 
M(\ld)=\left(\begin{array}{cc} w&1\\\ld&0\end{array}\right)\   . 
\end{equation} 
We will not discuss here the details of the explicit map 
on the level of the local variables $v_n$, (see the comments in 
section 5), but only work on the level of the monodromy matrix. 
What is essential is that the discrete-time evolution implies the 
invariance under the map of the spectral curve 
\begin{eqnarray} \label{eq:sp-curve} 
\Gamma: \qquad  \mathcal{R}(\eta,\ld)=\det(T(\ld)-\eta)=0\   ,
\quad n=1,\ldots,P,
\end{eqnarray}
which defines an hyperelliptic curve of
genus $g=P-1$, the branch points of which being defined by the 
formula 
\begin{equation}\label{eq:branch} 
\ol{\eta}^2=R(\ld)=\sum_{j=0}^{2g+1} r_j\ld^j= 
r_{2g+1}\prod_{j=1}^{2g+1} (\ld-e_j)\   , 
\end{equation} 
with $\ol{\eta}=\eta-\frac{1}{2}(A(\ld)+D(\ld))$. $R(\ld)$ is 
the discriminant of the hyperelliptic curve, and its branch 
points $e_i$, $i=1,\ldots, 2g+1$ as well as their symmetric 
functions $r_i$ of order $i$ are invariant under the discrete-time  
map (\ref{eq:Tmap}). In principle the branch points, which 
correspond to the eigenvalues of the associated $N\times N$ 
tridiagonal matrix, can be complex. To establish under what 
conditions they are real-valued and simple requires detailed 
analysis which is beyond the scope of this note. 

As is well-known from the continuous theory, \cite{FM,AvM}, the roots 
$\mu_j$, $j=1,\dots,g$ of the polynomial $B(\ld)$, which incidentally 
correspond to the poles of the Baker-Akhiezer function, define the 
so-called {\it auxiliary spectrum}, and they contain the relevant 
information on the dynamics of the system under consideration.   
Let us, therefore, derive the discrete equations for the auxiliary 
spectrum which then form the natural analogue 
of Dubrovin equations (\ref{eq:cDubr}). This will, in fact, lead to a
universal system of coupled first-order difference equations 
associated with the hyperelliptic  curve and to derive them we only 
need some global properties of the monodromy matrix. 

For genus $g$ the monodromy matrixes $T(\ld)$ takes on either one 
of two distinct possible forms depending on whether the length $N$ 
of the chain in (\ref{eq:T}) is even or odd. 
Writing eq. (\ref{eq:T}) as (in which we have taken $\ar_N=0$ 
in accordance with (\ref{eq:Tmap})), 
\begin{equation} \label{mono} 
T(\ld) = \left( \begin{array}{cc} 
\ld^{g+1}A_{g+1}+\ld^gA_g+\cdots + A_0 & \ld^gB_g+\ld^{g-1}B_{g-1}+ 
\cdots +B_0\\ 
\ld\left( \ld^gC_g+\ld^{g-1}C_{g-1}+ \cdots +C_0\right) & 
\ld\left(\ld^gD_g+\ld^{g-1}D_{g-1}+ \cdots +D_0\right)  
\end{array} \right)  \end{equation} 
the two cases associated with genus $g$ are either of the 
folowing two: 
\begin{equation} \label{eq:evenodd} {\rm Odd\ \ Case:} ~~~  
\left\{ \begin{array}{ccc} 
A_{g+1}=0&,&B_g=1\\ C_g=1&,&D_g=0 \end{array}\right\} 
\     \ ,\    \ {\rm Even\ \ Case:} ~~~  
\left\{ \begin{array}{ccc} 
A_{g+1}=1&,&B_g\neq 0 \\ C_g\neq 0&,&D_g=1 \end{array}\right\}\  .  
\end{equation} 
The even/odd distinction 
is related to the variance in the choice of periodic initial data 
in the two-dimensional lattice described by lattice equations of 
KdV type, cf. \cite{PNC}, where we can impose periodicity on 
chains (``staircases'') with period $2P-1$ respectively 
period $2P$ both corresponding to a curve of genus $g=P-1$. 
We note in passing that as a function of the invariants 
$I_j$, $j=0,\dots,g$ of the map which are the coefficients of 
the trace of the monodromy matrix 
\begin{equation}\label{eq:trace}
{\rm tr}\ T(\ld)=I_0+\sum_{j=1}^g I_j\ld^j\      \ ,\      \ 
I_j=A_j+D_{j-1}\  , 
\end{equation} 
the top coefficient $I_g=A_g+D_{g-1}$ being a Casimir with 
respect to the natural Poisson algebra associated with the dynamical 
map, the discriminant of the curve takes on the form 
\begin{equation}\label{eq:discr}
R(\ld)=\frac{1}{4}\left(I_0+\sum_{j=1}^gI_j\ld^j\right)^2-
\prod_{n=1}^N (-\ld_n)=B_gC_g\ld^{2g+1}+\ldots\  , 
\end{equation} 
thus depending on the shift variables $\ar_n$ entering via 
$\ld_n=\ld+\ar_n$. 

{}From the discrete-time map (\ref{eq:Tmap}) 
we have the following discrete relations for its entries 
\bea  
\wt{A}(\ld)=wB(\ld)+ D(\ld)\     \ &,& \    \ 
\wt{C}(\ld)=\ld B(\ld)\    \nn \\ 
\ld \wt{B}(\ld) = C(\ld) + w(A(\ld)-D(\ld)) -w^2B(\ld)\    \ &,& 
\   \ \wt{D}(\ld) =A(\ld)-wB(\ld)\  . \label{eq:ABCD} 
\eea 
Expanding eq. (\ref{eq:ABCD}) in powers of $\ld$ we are lead to the 
following set of equations: 
\begin{equation} \wt{A}_0 = wB_0=I_0\   \ ,\   \ \wt{A}_j=wB_j+D_{j-1}\   \ ,\   \ 
j=1,\dots,g\   . \label{eq:AuB} 
\end{equation} 
The entry $B(\ld)$ has the following factorisation:
\begin{equation}\label{eq:B}
B(\ld) = B_g\prod_{j=1}^g (\ld-\mu_j)\    , 
\end{equation} 
leading to the expressions for $B_0/B_g,\dots,B_{g-1}/B_g$ as 
elementary symmetric functions of the zeroes $\mu_1,\dots,\mu_g$.  
Similarly, the coefficients of $C(\ld)$ are symmetric functions of 
its zeroes $\undertilde{\mu_1},\dots,\undertilde{\mu_g}$, where 
the undertilde denotes the backward time-shift, and from (\ref{eq:ABCD}) 
we establish that the top coefficients $B_g$ and $C_g$ can be taken to 
be equal and constant. From the fact that 
\begin{equation} \label{eq:R} 
\frac{1}{2}(A-D)(\ld)=\kp\sqrt{R(\ld)-B(\ld)C(\ld)}\   ,   
\end{equation} 
where the $\kappa$ denotes the sign $\kp=\pm$ corresponding to the choice 
of sheet of the Riemann surface, subject to the condition 
$\wt{\kp}=-\kp$ (as follows from (\ref{eq:R}) and the relations 
(\ref{eq:ABCD})), and taking $\ld=\mu_i$ in (\ref{eq:R}) we obtain the 
system 
\begin{equation}\label{eq:system} 
\left( \begin{array}{c} 
\hf I_1 - D_0 \\ \vdots \\ \hf I_g -D_{g-1} \end{array}\right)  = 
\left(\begin{array}{ccc}
\mu_1 & \cdots & \mu_1^g \\ 
\vdots & & \vdots \\ 
\mu_g & \cdots &\mu_g^g  
\end{array} \right)^{-1}\cdot 
\left( \begin{array}{c} 
\kp\sqrt{R(\mu_1)}-\hf I_0 \\ \vdots \\ 
\kp\sqrt{R(\mu_g)}-\hf I_0 \end{array} \right)\  .  
\end{equation} 
On the one hand, (\ref{eq:system}) provides us with the values 
of $I_j-D_{j-1}-\wt{D}_{j-1}=wB_j$, ($j=1,\dots,g$), whereas 
on the other hand from the first of eq. (\ref{eq:AuB}) we note 
that $w=I_0/B_0$, hence we need the expressions for the factors 
$B_j/B_0$ which can be expressed in terms of symmetric functions 
of the $\mu_j$ as a consequence of (\ref{eq:B}). Alternatively, 
we can take $\ld=\undertilde{\mu_i}$ in (\ref{eq:R}) and thus 
obtain set of first-order difference equations for the $\mu_i$, 
namely 
\begin{equation}\label{eq:dDubr}
{\cal M}^{-1}\cdot\left(\kp\sqrt{R(\boldsymbol{\mu})}+
\hf I_0{\boldsymbol e}\right) + 
\wt{\cal M}^{-1}\cdot \left(\wt{\kp}\sqrt{R(\wt{\boldsymbol{\mu}})}-
\hf I_0{\boldsymbol e} \right) = 0 \   .   
\end{equation} 
In eq. (\ref{eq:dDubr}) $\boldsymbol{\mu}=
\left(\mu_1,\dots,\mu_g\right)^t$ denotes the vector with entries 
$\mu_j$ and ${\boldsymbol e}=(1,1,\dots,1)^t$, 
wheras $\mathcal{M}$ denotes the VanderMonde matrix 
\[  \mathcal{M}= \left(\begin{array}{ccc}
\mu_1 & \cdots & \mu_1^g \\ 
\vdots & & \vdots \\ 
\mu_g & \cdots &\mu_g^g  
\end{array} \right)\   .  \] 
We note that the terms with $I_0$ can be expressed in terms of 
the symmetric functions of the $\mu$'s and $\wt{\mu}$'s, namely 
by 
\[ {\cal M}^{-1}\cdot \boldsymbol{e}= - 
{\boldsymbol S}\left(-\mu_1^{-1},\dots,-\mu_g^{-1} \right) \  , \] 
where ${\boldsymbol S}(x_1,\dots,x_n)$ is the vector of symmetric 
products $S_k$ of its arguments, i.e. 
\[ S_k(x_1,\dots,x_g)\equiv \sum_{i_1<i_2<\dots<i_k} 
x_{i_1}x_{i_2}\dots x_{i_k}\    . \] 
Furthermore, in the spirit of \cite{Kent} we can obtain 
from (\ref{eq:ABCD}) and (\ref{eq:R}) by using an asymptotic 
expansion as ~$\ld\rightarrow\infty$~ a 
reconstruction formulae for the variable $w$, namely 
\begin{equation}\label{eq:wmu} 
w=\kappa\left( \sqrt{\cal A} - \sqrt{\wt{\cal A}}\right) 
\quad ,\quad {\cal A}\equiv  
\sum_{j=1}^g (\mu_j+\undertilde{\mu_j}) 
-\sum_{j=1}^{2g+1} e_j 
\end{equation} 
in terms of symmetric functions of the auxiliary spectrum. 
The ambiguity in the choice of sign reflects the 
reversibility of the map in the forward/backward discrete-time 
direction, and can be fixed at our choice subject to the condition 
that $\wt{\kappa}=-\kappa$. 

The discrete Dubrovin eqs. (\ref{eq:dDubr}) is a set 
of $g$ first-order difference equations for the $\mu_j$ 
and similarly as in the continuous case (\ref{eq:cDubr}) depending 
on the invariants $I_j$ only. Discrete Dubrovin equations were 
mentioned first in \cite{DMN}, but not given in explicit form. 
In \cite{Kent} they were interpreted as special 
addition formulae for Abelian functions. From the rather general 
and elementary derivation given above we conclude that these 
equations ar quite universal and linked to the Jacobi inversion 
problem on the hyperelliptic curve in a fundamental way. It 
would be of interest, therefore, to find a mechanism similar to 
the Lagrange interpolation trick that works in the continuous 
case to connect the discrete Dubrovin equations to linear 
motion on the Jacobian of the curve. So far, we have not been 
able to find such a ``direct'' integration, although some of the 
original works by Abel, e.g. \cite{Abel}, suggest that such a 
mechanism might exist. We have for the time being to resort to 
an indirect approach to linearise eqs (\ref{eq:dDubr}) which 
is by exploiting what is effectively the interpolating flows of 
the corresponding discrete-time maps. This will be discussed in 
the next section. 

We mention at this point also an 
alternative form of the discrete Dubrovin equations, namely as 
a second-order implicit system of equations of the form 
\begin{equation}\label{eq:altDubr} 
\wt{B}(\mu_i)-\undertilde{B}(\mu_i) = 2\kp\frac{w}{\mu_i}
\sqrt{R(\mu_i)}\   , \qquad i=1,\ldots,g\  . 
\end{equation}
This form of the discrete Dubrovin equations, which is closer to 
the forms proposed in \cite{DMN}, was the one discussed in 
\cite{Kent}. 

\section{Explicit Examples: $\boldsymbol{g=1}$ 
and  $\boldsymbol{g=2}$ } 
\setcounter{equation}{0}

Let us investigate the form of the discrete Dubrovin equations 
(\ref{eq:dDubr}) in the special cases of genus $g=1$ and $g=2$. 
In the case of $g=1$ the discrete Dubrovin 
equations (\ref{eq:dDubr}) reduce to one single equation, namely 
\begin{equation}\label{eq:dDubr1} 
\hf I_0\left( \frac{1}{\wt{\mu}} - \frac{1}{\mu}\right) 
= \frac{1}{\mu} \sqrt{R(\mu)} + \frac{1}{\wt{\mu}}
\sqrt{R(\wt{\mu})}\  , 
\end{equation} 
where for simplicity we have omitted the sign designation $\kp$. 
It is not hard to see that if we properly normalise the corresponding 
elliptic curve, i.e. (\ref{eq:sp-curve}) for $g=1$, eq. 
(\ref{eq:dDubr1}) is actually nothing else than the addition formula 
for the Weierstrass elliptic $\wp$-function, namely by identifying 
\bse\label{eq:Pg1}   
\begin{equation}
\mu=\wp(\dd) -\wp(t)\   \ , \   \ \wt{\mu}=\wp(\dd) 
-\wp(t+\dd)\   , 
\end{equation}  
together with 
\begin{equation} \ol{\eta}=\sqrt{R(\mu)} =\wp^\prime(t)\   \ ,\    \ 
-\hf I_0=\wp^\prime(\dd)\   , 
\end{equation}  \ese 
the parameter $\dd$ playing the role of the discrete time-step. 
Thus, in this case the discrete Dubrovin equation forms a natural 
difference equation for the Weierstrass elliptic $\wp$-function. 
Generic difference equations for elliptic functions were 
proposed by Potts \cite{Potts} a decade ago. 

In the case of genus $g=2$ the discrete Dubrovin equations 
are given by the following set of two coupled first-order 
difference equations: 
\bse \label{eq:dDubr2} \bea 
\hf I_0\left( \frac{1}{\mu_1\mu_2} - 
\frac{1}{\wt{\mu}_1\wt{\mu}_2}\right) 
&=& \frac{\frac{1}{\mu_1}\sqrt{R(\mu_1)} - 
\frac{1}{\mu_2}\sqrt{R(\mu_2)}}{\mu_1 -\mu_2} + 
\frac{\frac{1}{\wt{\mu}_1}\sqrt{R(\wt{\mu}_1)} - 
\frac{1}{\wt{\mu}_2}\sqrt{R(\wt{\mu}_2)}}{\wt{\mu}_1 -\wt{\mu}_2}
\    ,  \nn  \\ \\
\hf I_0\left( \frac{\mu_1+\mu_2}{\mu_1\mu_2} - 
\frac{\wt{\mu}_1+\wt{\mu}_2}{\wt{\mu}_1\wt{\mu}_2}\right) 
&=& \frac{\frac{\mu_2}{\mu_1}\sqrt{R(\mu_1)} - 
\frac{\mu_1}{\mu_2}\sqrt{R(\mu_2)}}{\mu_1 -\mu_2} + 
\frac{\frac{\wt{\mu}_2}{\wt{\mu}_1}\sqrt{R(\wt{\mu}_1)} - 
\frac{\wt{\mu}_1}{\wt{\mu}_2}\sqrt{R(\wt{\mu}_2)}}
{\wt{\mu}_1 -\wt{\mu}_2} \    .  \nn \\ 
 \eea \ese 
This system can be resolved along the following lines. 
Without giving any details of the analysis we mention that the $\mu_i$ 
arise as the poles, and the $\wt{\mu}_i$ as the zeroes of a 
``transition factor'' $\wt{\vf}/\vf$ where $\vf$ is the relevant 
Baker-Akhiezer function. In this case the independent variable 
$\boldsymbol{t}$ lives on the Jacobian $Jac(\Gamma)$ of the curve and 
we have the Jacobi inversion problem in the form: 
\begin{equation}\label{eq:t}  
\boldsymbol{t} = 
\int_{e_1}^{\mu_1} d\boldsymbol{u} + 
\int_{e_2}^{\mu_2} d\boldsymbol{u} \quad ,\quad 
d\boldsymbol{u}=\frac{\boldsymbol{a}+\mu\boldsymbol{b}}
{\sqrt{R(\mu)}} d\mu\   ,   
\end{equation} 
$d\boldsymbol{u}$ being a properly chosen vector (i.e. with 
appropriately chosen vectors $\boldsymbol{a}$ and $\boldsymbol{b}$) 
of holomorphic differentials on the curve, and $e_1$,$e_2$ being two 
of the branch 
points. Using Abel's theorem the map $\mu_i\rightarrow\wt{\mu}_i$ 
is resolved via the shifts on the Jacobian of the form 
\begin{equation} \label{eq:Abel} 
\boldsymbol{\dd} = \sum_{i=1,2} \int_{\mu_i}^{\wt{\mu}_i} 
d\boldsymbol{u} = 
-\int_\infty^0 d\boldsymbol{u} 
\quad \in \ \ Jac(\Gamma) 
\end{equation} 
in terms of the special winding vector (on the right hand side) on 
the Jacobian. 

In \cite{Kent} we have used Kleinian functions to parametrise 
the solution in terms of the genus $g=2$ analogue of the Weierstrass 
$\wp$-function, namely in terms of the Kleinian functions $\wp_{ij}$, 
$i,j=1,2$. 
The definitions of the generalised Weierstrass functions go 
back to Klein, cf. \cite{Klein}, and have also been discussed at 
great length in the monographs by Baker, \cite{Baker1,Baker2}. 
The general theory has been revived in the papers \cite{BEL}. 
(For an outline of the relevant definitions, cf. the Appendix). 
We restrict ourselves here to list the relevant formulae for the 
case $g=2$ to parametrise the solutions 
$\boldsymbol{\mu}=(\mu_1,\mu_2)$ of the discrete Dubrovin 
equations (\ref{eq:dDubr2}), are the following, cf. \cite{Baker2}:  
\bse \label{eqPg2} 
\begin{equation}
\wp_{22}(\bu)=\mu_1+\mu_2,\quad\wp_{12}(\bu)
=-\mu_1\mu_2\label{p},\quad
\wp_{11}(\bu)=\frac{F(\mu_1,\mu_2)-2\ol{\eta}_1\ol{\eta}_2}
{4(\mu_1-\mu_2)^2},\end{equation} 
where
\begin{equation}
F(\mu_1,\mu_2)=\sum_{k=0}^{2}\mu_1^k\mu_2^k(2r_{2k}
+ r_{2k+1}(\mu_1+\mu_2))\label{ff}\   , 
\end{equation} \ese 
in terms of the coefficients of the $g=2$ curve, cf. eq. 
(\ref{eq:sp-curve}).We note that only two of the functions 
$\wp_{ij}(\bu)$ are independent.
The three functions
$\wp_{22}(\bu)$, $\wp_{21}(\bu)$  and $\wp_{11}(\bu)$ are connected 
by a quartic relation which is 
the remarkable Kummer surface ${\mathcal K}$ in ${\mathbb C}^3$,
the equation of which is given by 
\begin{equation}\label{eq:kummer} {\mathcal K}:\quad 
\det K=\det\left(
\begin{array}{cccc}
-r_0&\frac12 r_1&2\wp_{11}&-2\wp_{12}\\
\frac12 r_1&-r_2-4\wp_{11}&\frac12 r_3+2\wp_{12}&2\wp_{22}\\
2\wp_{11}&\frac12 r_3+2\wp_{12}&-r_{4}-4\wp_{22}&2\\
-2\wp_{12}&2\wp_{22}&2&0
\end{array}
\right) = 0\   .  
\end{equation} 
The corresponding evaluations on the curve can be expressed 
in terms of the Kleinian three-indexed symbols, \cite{Baker2,Baker1}, 
by the relations 
\begin{equation} 
\ol{\eta}_1=\mu_1\wp_{222}(\bu)+\wp_{221}(\bu),\quad 
\ol{\eta}_2=\mu_2\wp_{222}(\bu)+\wp_{221}(\bu)\  .  
\label{3index} \end{equation} 
where these odd Kleinian functions are related to the even 
functions via 
\bse\bea 
\wp_{222}^2&=&4\wp_{11}+r_3\wp_{22}+4\wp_{22}^3+4\wp_{12}\wp_{22}
+r_4\wp_{22}^2+r_2\   ,\cr
\wp_{221}^2&=&r_0-4\wp_{11}\wp_{12}+r_4\wp_{12}^2
+4\wp_{22}\wp_{12}^2\   . 
\eea\ese 
What is needed still is to identify the arguments of the Kleinian 
functions with the variable $\bt$ on the Jacobian. This is rather 
subtle and will be discussed separately, cf. \cite{EN}.

\section{Connection with Separation of Variables}
\setcounter{equation}{0}

As has been mentioned earlier, the auxiliary spectrum $\{ \mu_j\}$ 
are on the one hand the poles of the properly normalised 
Baker-Akhiezer function, on the other hand they can be considered to 
be separation variables for the integrable Hamiltonian systems, 
cf. \cite{FM,AvM}. 
In recent papers, \cite{Skly,AHH}, this connection has been 
put into a bigger framework to understand the general problem 
of seperation of variables for integrable systems going 
beyond the classical coordinate separation. The general statement 
is the following, \cite{Skly}:\\ 
{\it For a finite-dimensional integrable Hamiltonian system 
$(\{q_i\},\{p_i\},H)$ admitting a Lax 
representation, i.e. for which a matrix $L(\ld)$ exists such that 
the spectral curve 
\[ \Gamma:\      \  \det\left( L(\ld)-\eta\right)=0  \] 
yields a complete set of integrals of the motion in involution, the 
poles 
$\{\mu_i\}$ of the Baker-Akhiezer function, i.e. the eigenfunction 
of the Lax matrix when properly normalised, together with the 
corresponding evaluations on the curve, i.e. $\eta_i=\eta(\mu_i)$ 
form a set 
of separation variables for the Hamiltonian system, i.e. 
there is a canonical transformation 
\[ (q_i,p_i)\  \ \rightarrow\   \ (\mu_i,\eta_i)   \] 
and the separation equations are given by  
\[ \det\left( L(\mu_i)-\eta_i\right) =0   \] 
expressing the fact that the $(\mu_i,\eta_i)$ are lying on the 
spectral curve. } 
\\ 
This statement has been exemplified by a large number of 
examples, cf. \cite{Skly}, and notably recently by generic 
systems of Calogero-Moser and Ruijsenaars type even in the full 
elliptic case, \cite{KNS}. We point out that the question of the 
normalisation of the Baker-Akhiezer function is a crucial point 
and no general prescription exists to date.  

The separation mechanism is rather independent of the type of 
dynamics we impose on the Hamiltonian system. Thus, it applies 
equally well to situations where we have an integrable  discrete-time 
dynamics described by a Lax pair, in which case the corresponding 
discrete-time map is an (iterative) canonical transformation 
which usually can be cast into a discrete Lagrangian form, 
cf. e.g. \cite{Ves}. This applies to the mappings of the type studied 
in the present paper, cf. \cite{CNP}, and in \cite{NPC} for mappings 
of this type a classical $r$-matrix structure was found from which 
we can extract the following Poisson brackets: 
\bse\label{eq:Pbrackets} \bea 
\{ A(\ld)\,,\,B(\ld')\,\} &=& \frac{B(\ld)A(\ld')-A(\ld)B(\ld')}
{\ld-\ld'} + \frac{1}{\ld'}B(\ld)D(\ld')\    \\ 
\{\,B(\ld)\,,\,B(\ld'\,\} &=& 0    \\ 
\{ D(\ld)\,,\,B(\ld)\,\} &=& \frac{1}{\ld'}   
\frac{\ld'D(\ld)B(\ld')-\ld B(\ld)D(\ld')}
{\ld-\ld'} 
\eea\ese   
Eqs. (\ref{eq:Pbrackets}) can be used to establish the canonicity 
of the separation variables, i.e. 
\begin{equation}\label{eq:canon} 
\{ \eta_i\,,\,\eta_j \} = \{ \mu_i\,,\,\mu_j \} = 0\      \ ,
\       \ \{ \eta_i\,,\,\mu_j \} = \dd_{ij}\eta_i\    , 
\end{equation} 
using ~$\eta(\mu_i)=A(\mu_i)$~. 
The symplecticity of the discrete-time map is guaranteed from the 
Poisson brackets involving the variable $w$: 
\bea \label{eq:Tsympl} 
\{ A(\ld)\,,\,w \} &=&\frac{1}{\ld}\left( wA(\ld)-w^2B(\ld)-C(\ld)
\right) \      \ ,\      \ 
\{ C(\ld)\,,\,w \} =\frac{w}{\ld}\left( C(\ld)-wD(\ld) \right) 
\nn  \\ 
\{ B(\ld)\,,\,w \} 
&=& \frac{1}{\ld}\left( A(\ld)-D(\ld)-wB(\ld) \right) 
\      \ ,\       \ 
\{ D(\ld)\,,\,w \} =\frac{1}{\ld}\left( C(\ld)-wD(\ld) \right) \  . 
\eea
Thus, the general separation mechanism is consistent with the 
discrete-time evolution. 

Let us finish with some general remarks. It seems that the 
discrete-time systems and integrable mappings play a more profound 
role in the general problem of separation, cf. \cite{KS}. In 
principle, the 
problem of finding the separation of variables in terms of 
discrete Dubrovin equations of the type (\ref{eq:dDubr}), which 
for a $N$-dimensional Hamiltonian system generically take the form 
\begin{equation}\label{eq:gDubr}
G_i(\{\wt{\mu}_i\},\{\mu_i\};I_1,\dots,I_N)=0\      \ ,\     \ 
i=1,\dots,N 
\end{equation}
can be formulated as a problem of finding {\em commuting canonical 
transformations}. Thus, we might conjecture that in the 
relevant cases the following diagram actually represents a 
commuting diagram of integrable canonical maps as follows:  

\begin{picture}(100,120)  
\put(30,80){\makebox[2cm][r]{$(q_i,p_i)$}  }
\put(30,20){\makebox[2cm][r]{$(\mu_i,\eta_i)$}  } 
\put(170,80){\makebox[2cm][r]{$(\wt{q}_i,\wt{p}_i)$}  }
\put(170,20){\makebox[2cm][r]{$(\wt{\mu}_i,\wt{\eta}_i)$}  } 
\put(80,80){\makebox[4cm]{$\stackrel{\textstyle S(q_i,\wt{q}_i)}
 {\longrightarrow}$ } } 
\put(10,50){\makebox[3cm]{$F(q_i,\mu_i)\   \ 
\downarrow$} } 
\put(190,50){\makebox[3cm]{$\downarrow\   \ F(\wt{q}_i,\wt{\mu}_i)$} } 
\put(80,10){ \makebox[4cm]{$\stackrel{\longrightarrow}
 {W(\mu_i,\wt{\mu}_i)}$} } 
\end{picture} 

In this diagram the $S$ is the action functional describing the 
canonical transformation which is the discrete mapping (as can be 
derived from a discrete Lagrangian), and $F$ denotes the 
generating function of the canonical transformation from the 
original variables to the separating variables. The canonical 
transformation, with generating function $W$, realising
the dynamical mapping in terms of the separation variables, is 
obviously, by construction, an integrable map itself. Unfortunately, 
however, it does not seem easy in general to obtain explicit 
expressions for the generating function $W$, since this requires 
the elimination of the invariants entering as coefficients of the 
spectral curve. Noting that the general issue of superposition of 
canonical transformations and the connected problem of finding 
commuting canonical maps has to our knowledge not been addressed 
in full generality in classical mechanics, it seems that the advent 
of integrable discrete dynamical systems will make the study of 
these questions imminent.

\section{Conclusions} 
\setcounter{equation}{0}

In \cite{PNC} we constructed a family of exactly integrable 
finite-dimensional mappings  from a lattice KdV equation 
by considering `local' initial value problems on 
so-called `staircases' in the lattice. These mappings are symplectic 
as consequence of the Lagrange structure of the original lattice 
KdV equation, and their complete integrability in the sense 
of Liouville, \cite{Ves}, was established in \cite{CNP,NPC}. The 
resulting mappings can be written in the form of multidimensional 
rational mappings ${\mathbb R}^N\rightarrow {\mathbb R}^N:
\{v_n\}\mapsto\{\wt{v}_n\}$, $n=1,\dots,N$, where the variables 
$v_n$ coincide with the ones entering in the monodromy matrix 
(\ref{eq:T}). In fact, their Lax 
description is based on the matrices $V_n$ used in section 2.  
In the recent paper, \cite{Kent}, V. Enolskii and the author 
investigated the finite-gap integration of the resulting mappings 
and their parametrisation in terms of the Kleinian functions. 
It was pointed out there that these integrable mappings have, 
in fact, the interpretation of being addition formulae for 
hyperelliptic Abelian functions for special winding vectors on 
the Riemann surface. In that paper we concentrated on the {\it 
local description} of the mappings, taking into account the 
dependence of the variables $\mu_i$ on the variable $n$ 
labelling the sites along the periodic chain as encoded in the 
Lax matrices $V_n$. Interestingly, the shift variables $\ar_n$ 
play an essential role in this description since they enter 
as distinct singularities on the curve defining special winding 
vectors corresponding to the shifts $n\mapsto n+1$. 
In the present paper we have not at all 
addressed that issue, nor the actual reconstruction formulae 
for the potential $v_n$. We should also mention the 
work by Bobenko and Pinkall, \cite{BP}, on the geometric 
aspects of lattice systems associated with equations of KdV type, 
where from a different perspective finite-gap formulae have also 
been presented.

\section*{Acknowledgement} 

Part of this work has been performed in collaboration with 
Victor Enolskii and a more detailed account will appear elsewhere, 
\cite{EN}. The author acknowledges discussions with 
Vadim Kuznetsov and Evgueni Sklyanin.

\section*{Appendix: Kleinian Functions} 
\def\theequation{A.\arabic{equation}}
\setcounter{equation}{0}

In this appendix we collect a number of relevant formulae on 
Kleinian functions, \cite{Klein,Baker1,Baker2}. We refer the 
reader to the excellent review papers \cite{BEL}, (from which 
this material is taken), in which this 
theory has been cast in a modern context and where it was shown 
that these functions arise naturally within the KdV theory. 

Let the hyperelliptic curve be given by $\Gamma: y^2=R(x)$, 
as in (\ref{eq:sp-curve}), of genus 
$g$ having $2g+2$ branch points $e_1,\ldots,e_{2g+1},\infty$, 
cf. also (\ref{eq:branch}), and as is well-known we 
can equip it with a canonical homology basis  (${\mathfrak a}_1$,
\ldots, ${\mathfrak a}_g$; ${\mathfrak b}_1$,\ldots,${\mathfrak b}_g$), 
and define on $\Gamma$ a canonical set of holomorphic differentials,
\begin{equation}  
d{\bf u}=(du_1,\ldots,du_g),
\quad
du_k=\frac{x^{k-1} dx}{y}\   , 
\end{equation} 
and differentials of the second kind with a pole at infinity
\begin{equation} 
d{\boldsymbol \Omega}=(d\Omega_1,\ldots, d\Omega_g),\quad , \quad 
d\Omega_j= \sum_{k=j}^{2g-j} (k+1-j) r_{k+1+j} 
\frac{x^k dx }{4y}\  .   
\end{equation} 
Introducing $g\times 2g$ period matrices $(2\boa,2\boa^\prime)$ and
$(2\bet,2\bet^\prime)$ of their respective integrals over the 
${\mathfrak a}$-- and ${\mathfrak b}$ cycles, we note that 
$\det \omega\neq 0$ and the matrix 
$\boldsymbol{\tau}=\boa^{-1}\boa^\prime$ 
is symmetric and its imaginary part is positive definite. 
Introducing also the matrix $\varkappa= \bet(2\boa)^{-1}$, 
we can now define the fundamental hyperelliptic 
$\sigma$--function by the formula
\begin{equation}
\sigma({\boldsymbol u})={\sqrt\frac{\pi^g}{\det \;2\boa}}
\frac{1}{\sqrt[4]{\prod_{i\neq j}(e_i-e_j)}}
\mathrm{exp}\{{\boldsymbol u}^t{\varkappa }{\boldsymbol u}\}
\theta[\varepsilon]
((2\boa)^{-1}{\boldsymbol u}|{\boldsymbol \tau}),
\end{equation}
where  $[\varepsilon]=\left[\begin{array}{c}
\boldsymbol{{\varepsilon}^\prime}\\\boldsymbol{\varepsilon}
\end{array}\right]$,  is the characteristic of the vector of Riemann
constants,
and $\theta[\varepsilon](\boldsymbol{v}|{\boldsymbol \tau})$ is 
the Riemann theta function with characteristic,
\begin{equation}
\theta[\ven]({\boldsymbol v}|\boldsymbol{\tau})
=\sum_{{\bm}\in {\mathbb Z}^g}\mathrm {\exp}
\{ \pi \imath ({\bm}+ \boldsymbol{\ven}^\prime)^t 
{\boldsymbol \tau}({\bm}+ \boldsymbol{\varepsilon}^\prime)
+ 2\pi \imath ({\bm}+ \boldsymbol{\varepsilon}^\prime)^t
({\boldsymbol v}+ \boldsymbol{\varepsilon})\}\   .  
\end{equation}
The Kleinian $\wp$--functions are defined by the second-- and 
third order logarithmic derivatives of the Kleinian 
$\sigma$--function, namely 
\begin{equation}
\wp_{ij}(\boldsymbol{u})
=-\frac{\partial^2\; \ln\,\sigma(\boldsymbol{u})}{\partial u_i\partial
u_j}\;,\quad \wp_{ijk}(\boldsymbol{u})
=-\frac{\partial^3\;\ln\,\sigma(\boldsymbol{u})}{\partial u_i\partial
u_j\partial u_k}\;,\quad i,j,k=1,\ldots,g. 
\end{equation}
The Abel map $\mathfrak{A}:(\Gamma)^g\rightarrow\ Jac(\Gamma)$ of
the symmetrised product $\Gamma\times\cdots\times\Gamma$ to the 
Jacobi variety $Jac(\Gamma)={\mathbb C}^g/2\boa\oplus 2\boa^\prime$ 
of the curve $\Gamma$ is defined by the vector relation 
\bea 
\sum_{i=1}^g \int_{(0,e_i)}^{(y_i,x_i)} d\boldsymbol{u} 
= \boldsymbol{u}\  ,  
\eea 
and where the $e_i$, $i=1,\dots,g$  are a properly chosen selection of 
branch points among the $2g+2$ branch points (except $\infty$) on the curve.  

The Jacobi inversion problem of inverting the Abel map, in the 
formulation by Weierstrass, can be solved in terms of the Kleinian
functions as follows:  The Abel preimage of the point $
\boldsymbol{u}\in Jac(\Gamma) $ is given  by the set $
\{ (y_1, x_1) ,   \ldots,  (y_g, x_g)   \}  \in  (\Gamma)^g$,
where $
\{x_1,   \ldots, x_g  \}$ are the zeros of the polynomial
\begin{equation}\label{x} 
{\cal P} (x;  \boldsymbol{u})
=x^g-x^{g-1}\wp_{g,g} ( \boldsymbol{u})-x^{g-2}\wp_{g,g-1}
(\boldsymbol{u})-\ldots-\wp_{g,1}(\boldsymbol{u}) ,
\end{equation} 
and  $\{y_1,\ldots,y_g\}$ are given by 
\begin{equation} \label{y} 
y_k=-\frac{\partial {\cal P}
(x;\boldsymbol{u}) }{\partial u_g}\Bigl\lvert_{x=x_k}. \;
\end{equation}
More details can be found in \cite{BEL}.

\pagebreak

\end{document}